\documentclass[conference]{IEEEtran}

\usepackage{enumerate}

\usepackage{cite}
\usepackage{graphicx,epsfig,amssymb}

\ifCLASSINFOpdf
\else
\fi
\usepackage[cmex10]{amsmath}
\begin{document}
%
% paper title
% can use linebreaks \\ within to get better formatting as desired
\title{Performance Analysis of Urban Mmwave Multi-hop V2V Communications with Shifted-Exponential Distribution Headway}

% conference papers do not typically use \thanks and this command
% is locked out in conference mode. If really needed, such as for
% the acknowledgment of grants, issue a \IEEEoverridecommandlockouts
% after \documentclass

% for over three affiliations, or if they all won't fit within the width
% of the page, use this alternative format:
%
\author{\IEEEauthorblockN{Zipeng Li,
Yangwanli Guo,
Xiaohu Ge
}
\IEEEauthorblockA{School of Electronic Information and Communications, Huazhong University of Science and Technology, China}
%\IEEEauthorblockA{\IEEEauthorrefmark{2}Department of Computer Convergence Software, Korea University, South Korea}
\IEEEauthorblockA{Corresponding Author: Xiaohu Ge, Email: xhge@mail.hust.edu.cn}}

% use for special paper notices
%\IEEEspecialpapernotice{(Invited Paper)}

% make the title area
\maketitle

\begin{abstract}
%\boldmath
With the emergence of autonomous driving and 5G mobile communication systems, how to design vehicular networks to meet the requirements of ultra-reliable low-latency communication (URLLC) between autonomous vehicles is widely concerned by the academia and industry. In this paper, message transmission delay and reliability between autonomous vehicles based on millimeter wave (mmwave) multi-hop V2V communications under urban road environment are investigated. Considering the fact that vehicles cannot come arbitrarily close to each other, shifted-exponential distribution is proposed to model the headway distance (the distance between the head of a vehicle and the head of its follower). Simulation results show that message transmission delay and reliability decrease with the increase of the path loss exponent and are also affected by the minimum safe distance between adjacent vehicles.
\end{abstract}
\vspace{2 ex}


\IEEEpeerreviewmaketitle

\section{Introduction}

\label{sec1}
% no \IEEEPARstart

Autonomous driving is a new direction in the development of automobile industry, and its goal is to meet the high demand for driving safety, comfort and reliability \cite{3Ge17}. In order to achieve safe autonomous driving, ultra-reliable low-latency communication (URLLC) among vehicles should be guaranteed. Therefore, how to devise vehicular networks to meet the stringent latency and reliability requirements for autonomous vehicles is widely concerned by the academia and industry.

With the emergence of 5G mobile communication system \cite{4Kong17,GeYe16,GeTu16}, millimeter wave is considered to be the key technology to realize the communications among autonomous vehicles due to its large bandwidth and good directivity. However, the millimeter wave signal has the defects of large attenuation and poor penetration, which limits the transmission range of millimeter wave signal and makes it impossible to communicate directly between two vehicles which are far away from each other. In this case, two autonomous vehicles that are far away from each other can establish a multi-hop V2V link to realize message transmission by selecting some vehicles between them as relays.

There are some related work to investigate the message transmission between vehicles based on V2V communications \cite{5Li17,6Farooq,7Li15,8Jeyaraj17}. Considering the delay and link connection probability, S. Li et al. proposed a new link quality indicator to evaluate the quality of multi-hop V2V link \cite{5Li17}. Simulation results show that there is an optimal average single-hop distance for the best link quality. In \cite{6Farooq}, considering the Carrier Sense Multiple Access (CSMA) protocol, authors studied the multi-hop V2V link message transmission in multi-lane highway scenarios. In particular, the trade-off between the FP (Forward Progress) and the single-hop link transmission success probability with different relay selection strategies are investigated. In \cite{7Li15}, X. Li et al. studied security message broadcasting based on multi-hop V2V links. The influence of single-hop communication distance and road vehicle density on the delay of safety message broadcast is analyzed. In \cite{8Jeyaraj17}, the Poisson point process is used to model the position of vehicles on orthogonal roads, and the probability of successful transmission of typical general vehicles and vehicle messages at intersections are investigated. It is found that the interference signals are mainly from vehicles located on the same road.

However, in the above research work, the Poisson point process is used to model the positions of vehicles on the roads. In Poisson point process, the headway distance (the distance between the head of a vehicle and the head of its follower) is subjected to exponential distribution, which makes that vehicles can come arbitrarily close to each other. In the actual traffic environment, a minimum safe distance must be guaranteed between adjacent vehicles to ensure safe driving. Therefore, it is not reasonable to use the Poisson point process to model the positions of vehicles on the roads.

Motivated by the above gaps, shifted-exponential distribution is proposed to model the headway distance in this paper. In the shifted-exponential distribution, a minimum safe distance can be guaranteed between adjacent vehicles. On this basis, message transmission delay and reliability between autonomous vehicles based on millimeter wave multi-hop V2V communications with random relay selection strategy under urban road environment are investigated.

The reminder of this paper is organized as follows. Section II describes the system model. The average total delay and reliability of message transmission with random relay selection strategy are investigated in Section III. In Section IV, simulations results show the impacts of path loss exponent and minimum safe distance between adjacent vehicles on average total delay and reliability of message transmission. Section V concludes this paper.

% You must have at least 2 lines in the paragraph with the drop letter
% (should never be an issue)

% Note that IEEE does not put floats in the very first column - or typically
% anywhere on the first page for that matter. Also, in-text middle ("here")
% positioning is not used. Most IEEE journals/conferences use top floats
% exclusively. Note that, LaTeX2e, unlike IEEE journals/conferences, places
% footnotes above bottom floats. This can be corrected via the \fnbelowfloat
% command of the floats package.

\section{System Model}
\label{sec2}

\subsection{Network Model}
\label{sec2-1}

\begin{figure*}
  \centering
  % Requires \usepackage{graphicx}
  \includegraphics[width=15cm,draft=false]{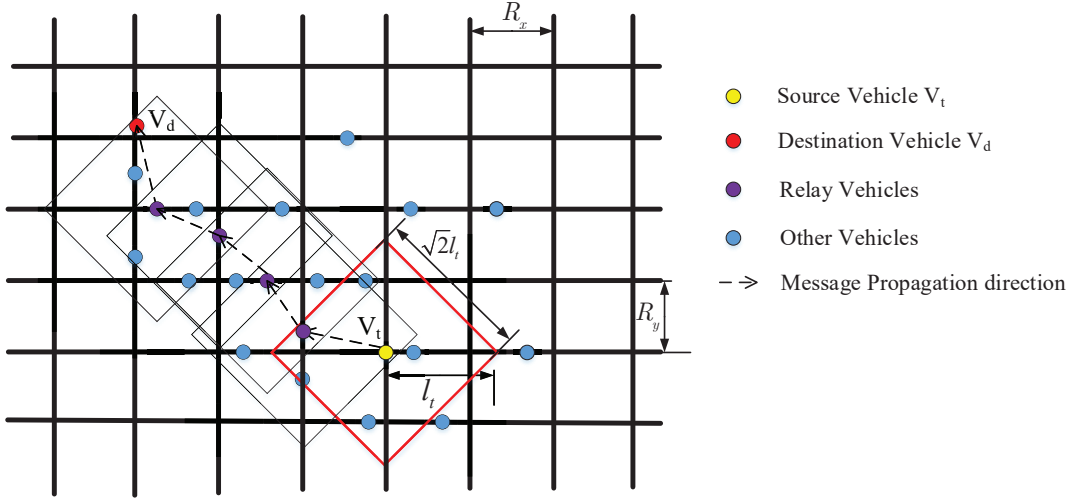}\\
  \caption{\small Network model.}
  \label{fig1}
\end{figure*}

Considering the network model shown in Fig. 1, let us ignore the width of roads and model the city roads as a grid. There are two types of roads: horizontal roads and vertical roads. The spacing between adjacent horizontal roads is equal to ${R_y}$, and the spacing between adjacent vertical roads is equal to ${R_x}$. The headway distance in each road is subjected to independent identically shifted exponential distribution. According to the definition of shifted exponential distribution \cite{9Cowan75}, the headway distance ${D_{{\rm{adj}}}}$ consists of two parts, i.e., ${D_{{\rm{adj}}}} = {d_{safe}} + U$. ${d_{safe}}$ is the tracking component, which is a constant and represents minimum safe distance that must be guaranteed between adjacent vehicles. U is the free component, which is subjected to exponential distribution with mean $\frac{1}{\mu }$.

As shown in Fig. 1, assuming that a traffic accident happens at an intersection. The typical autonomous vehicle ${V_t}$ (yellow node in Fig. 1) at the intersection knows this warning information, which needs to be sent to the autonomous vehicle ${V_d}$ (red node in Fig. 1) which intends to go to this intersection. The Euclidean distance between ${V_t}$ and ${V_d}$ is ${R_{valid}}$. Millimeter wave (mmwave) is adopted to the communications among all the vehicles and the transmitting power of each vehicle is a constant ${P_t}$. Since the path loss of millimeter wave signal is usually very severe, the effective transmission distance of millimeter wave signal is limited. Therefore, we assume that the effective communication distance of each vehicle is ${l_t}$, and ${l_t} < {R_{valid}}$. Since the mmwave signal is easily blocked by obstacles on both sides of the urban road, it is not easy to penetrate the obstacles. Therefore, it can be considered that the main energy of the mmwave signal propagates along the road. Hence the effective communication range of a single vehicle can be regarded as a circle defined in taxicab geometry (later called Manhattan circle) with the vehicle as the center and ${l_t}$ as the radius. The Manhattan circle with radius ${l_t}$ is a square with rotation of 45 degrees, and the side length of the square is $\sqrt 2 {l_t}$ \cite{10Krause73}. Taking the vehicle ${V_t}$ in Fig. 1 as an example, the effective communication range is a square with red edge length $\sqrt 2 {l_t}$. In this case, the propagation distance of the mmwave signal between the two vehicles can be characterized by the Manhattan distance between this two vehicles. Manhattan distance is defined as follows:

In the Cartesian coordinate system, the Manhattan distance between $\left( {{x_1},{y_1}} \right)$ and $\left( {{x_2},{y_2}} \right)$ is $\left| {{x_1} - {x_2}} \right| + \left| {{y_1} - {y_2}} \right|$.

Other vehicles within the effective communication range of a vehicle can communicate with the vehicle directly, while other vehicles outside the effective communication range cannot communicate with this vehicle directly. In order to ensure that messages can be transmitted from ${V_t}$ to ${V_d}$ successfully, some vehicles between ${V_t}$ and ${V_d}$ should be selected as relays to establish a multi-hop V2V link to complete the message transmission. Since the packet size of warning message is usually small and the latency requirement is typically millisecond, the positions of vehicles on the roads are supposed to remain unchanged during the period of message transmission. In order to achieve URLLC among vehicles, only one vehicle node in the multi-hop link is allowed to transmit at any moment. Therefore, there is no interference between vehicles. After the signal is processed in the relay vehicle (the noise is eliminated and the useful signal is extracted), the useful signal is sent to the next vehicle with transmitting power ${P_t}$ . Random relay selection strategy is adopted to select relay vehicles, i.e., the transmitting vehicle will select relay vehicle randomly within its effective communication range with positive forward progress (FP) \cite{6Farooq}.

\subsection{Sectorized Antenna Model}
\label{sec2-2}

To leverage array gain, directional beamforming by multiple antennas is performed at each vehicle. Transmission and reception directivity gains $g_{i,j}^\wp \left( {\wp  \in \left\{ {{t_x},{r_x}} \right\}} \right)$ of vehicles in link ${l_{i,j}}$ are given by \cite{11Perfecto17}

\begin{equation}
g_{i,j}^\wp  = \left\{ {\begin{array}{*{20}{c}}
  {{G_\sphericalangle },{\kern 1pt} {\kern 1pt} {\kern 1pt} {\kern 1pt} {\kern 1pt} {\kern 1pt} {\kern 1pt} }&{{\kern 1pt} if{\kern 1pt} {\kern 1pt} \left| {\alpha _{i,j}^\wp } \right| \leqslant {{\varphi _{i,j}^\wp } \mathord{\left/
 {\vphantom {{\varphi _{i,j}^\wp } 2}} \right.
 \kern-\nulldelimiterspace} 2}} \\
  {{g_\sphericalangle },}&{otherwise}
\end{array}} \right.,
\end{equation}
where ${G_\sphericalangle }$ denotes the main lobe gain, ${g_\sphericalangle }$ denotes the side lobe gain, $\alpha _{i,j}^\wp $ denotes the alignment error of antenna steering directions between transmitter ${V_i}$ and receiver ${V_j}$ and $\varphi _{i,j}^\wp $ denotes beam-level beamwidth.

\subsection{Channel Model}
\label{sec2-3}

72 GHz mmwave channel model is adopted in this paper \cite{12Ghosh14}. The channel gain $g_{i,j}^{\rm{c}}[dB]$ on a typical single hop link ${l_{i,j}}$ can be written as

\begin{equation}
g_{i,j}^{\text{c}}[dB]\left( {{d_{man,i - j}}} \right) = 69.6{\text{ + }}10\alpha {\log _{10}}\left( {{d_{man,i - j}}} \right) + \rho ,
\end{equation}
where $\rho $ is the shadow fading coefficient, $\rho  \sim N\left( {0,{\sigma ^2}} \right)$ and $\sigma $ is the standard deviation of shadow fading; $\alpha $ is the path loss exponent; ${d_{man,i - j}}$ is the Manhattan distance between ${V_i}$ and ${V_j}$.

\subsection{Alignment Delay and Transmission Rate}
\label{sec2-4}

There is a key technical problem in the application of millimeter wave in V2V communication, that is, how to achieve beam alignment. Therefore, we design the beam steering according to the following mechanism: before the transmission of effective data, the transmitter continuously transmits pilot data, and the receiver gives feedback to find the best beam steering. A two-staged beam alignment is adopted. These two stages encompass a sequence of pilot transmissions and use a trial-and-error approach where first a coarse sector-level scan detects best sectors for transmitter and receiver and, afterwards, within the limits of the selected sector a finer granularity beam-level sweep searches for best beam-level pairs. Without loss of generality, we assume here that for each vehicle in a V2V link before the phase of beam-level alignment, the sector-level alignment has already been performed. Then only the beam-level alignment should be done. Adopting the exhaustive search strategy, alignment delay in link ${l_{i,j}}$ is given by

\begin{equation}
{\tau _{i,j}} = \frac{{\psi _{i,j}^{{t_x}}\psi _{i,j}^{{r_x}}}}{{\varphi _{i,j}^{{t_x}}\varphi _{i,j}^{{r_x}}}}{T_p},
\end{equation}
where $\psi _{i,j}^{{t_x}}$ and $\psi _{i,j}^{{r_x}}$ denote the sector-level beamwidths of transmitter ${V_i}$ and receiver ${V_j}$, respectively, $\varphi _{i,j}^{{t_x}}$ and $\varphi _{i,j}^{{r_x}}$ denote the beam-level beamwidths of ${V_i}$ and ${V_j}$, respectively, and ${T_p}$ denotes the pilot transmission duration.

Therefore, the transmission rate ${r_{i,j}}$ of effective data in the typical link ${l_{i,j}}$ during the period of ${T_t}$ can be written as

\begin{equation}
\begin{gathered}
  {r_{i,j}} = \left( {1 - \frac{{{\tau _{i,j}}}}{{{T_t}}}} \right)B{\log _2}\left( {1 + SN{R_j}} \right) \hfill \\
   = \left( {1 - \frac{{{\tau _{i,j}}}}{{{T_t}}}} \right)B{\log _2}\left( {1 + \frac{{{P_t}g_{i,j}^{{t_x}}g_{i,j}^cg_{i,j}^{{r_x}}}}{{{N_0}B}}} \right) \hfill \\
\end{gathered},
\end{equation}
where $SN{R_j}$ is the signal-noise ratio, ${N_0}$ is the Gaussian background noise power density, and $B$ is the bandwidth of the mmwave band.

\subsection{The Average Total Delay of Message Transmission}
\label{sec2-5}

Assuming the packet size of warning message to be transmitted is ${P_S}$, the message transmission delay over single hop link ${l_{i,j}}$ is given by

\begin{equation}
{T_{i,j}} = \frac{{{P_S}}}{{{r_{i,j}}}} = \frac{{{P_S}}}{{\left( {1 - \frac{{{\tau _{i,j}}}}{{{T_t}}}} \right)B{{\log }_2}\left( {1 + SN{R_j}} \right)}}.
\end{equation}

The average total delay of message transmission over multi-hop V2V link is related not only to the transmission delay of message over the single hop link, but also to the hop number of the multi-hop V2V link. For the sake of simplicity, the number of hops in a multi-hop link is approximately given by

\begin{equation}
k \approx {{{R_{valid}}} \mathord{\left/
 {\vphantom {{{R_{valid}}} {\mathbb{E}\left( {{Z_{i,j}}} \right)}}} \right.
 \kern-\nulldelimiterspace} {\mathbb{E}\left( {{Z_{i,j}}} \right)}},
\end{equation}
where ${R_{valid}}$ is the Euclidean distance between source vehicle and destination vehicle, $\mathbb{E}\left(  \bullet  \right)$ is the expectation operator, and ${Z_{i,j}}$ is the FP of the typical link ${l_{i,j}}$.

Assuming the processing delay in each relay vehicle is ${T_{proc}}$, the average total delay of message transmission over multi-hop V2V link is derived by

\begin{equation}
\begin{gathered}
  {T_{multi}} = k\mathbb{E}\left( {{T_{i,j}}} \right) + \left( {k - 1} \right){T_{proc}} \hfill \\
   = k\mathbb{E}\left[ {\frac{{{P_S}}}{{\left( {1 - \frac{{{\tau _{i,j}}}}{{{T_t}}}} \right)B{{\log }_2}\left( {1 + SN{R_j}} \right)}}} \right] + \left( {k - 1} \right){T_{proc}} \hfill \\
   = k\int_0^{ + \infty } {\frac{{{f_{SN{R_j}}}\left( \gamma  \right){P_S}}}{{\left( {1 - \frac{{{\tau _{i,j}}}}{{{T_t}}}} \right)B{{\log }_2}\left( {1 + \gamma } \right)}}d\gamma }  + \left( {k - 1} \right){T_{proc}} \hfill \\
\end{gathered} ,
\end{equation}
where ${f_{SN{R_j}}}\left( \gamma  \right)$ is the PDF of $SN{R_j}$ over link ${l_{i,j}}$.

\subsection{The Average Total Reliability of Message Transmission}
\label{sec2-6}

The reliability over the typical single hop link ${l_{i,j}}$ is defined as the probability of $SN{R_j}$ larger than the threshold $\varepsilon $, which is given by

\begin{equation}
{P_{one,i - j}} = P\left( {SN{R_j} \geqslant \varepsilon } \right) = 1 - {F_{SN{R_j}}}\left( \varepsilon  \right),
\end{equation}
where ${F_{SN{R_j}}}\left( \varepsilon  \right)$ is the CDF of $SN{R_j}$.

Therefore, the average total reliability of message transmission over multi-hop V2V link is given by

\begin{equation}
{P_{multi}} = {\left( {{P_{one,i - j}}} \right)^k} = {\left[ {1 - {F_{SN{R_j}}}\left( \varepsilon  \right)} \right]^k}.
\end{equation}

\section{The Average Total Delay and Reliability of Message Transmission with Random Relay Selection Strategy}
\label{sec3}

For the sake of simplicity, assuming that there is a linear correspondence between the urban road length covered in a region and the area of this region \cite{13Fang15}, i.e., ${d_{area}} = \eta {S_{area}}$ , where ${d_{area}}$ is the road length covered in a region and ${S_{area}}$ is the area of this region.

\begin{figure}
  \centering
  % Requires \usepackage{graphicx}
  \includegraphics[width=9.5cm,draft=false]{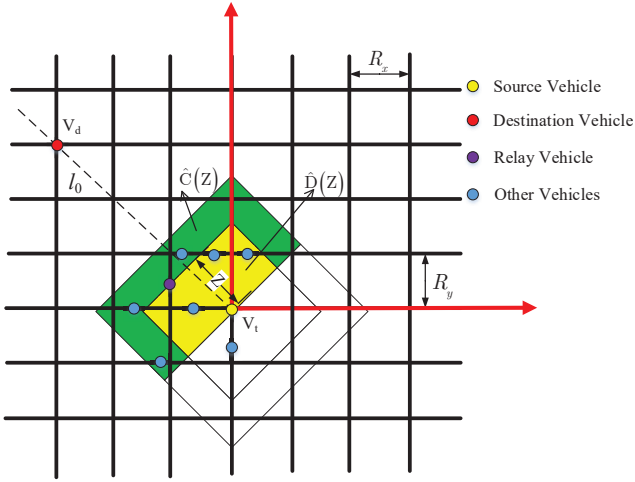}\\
  \caption{\small Random relay selection strategy.}
  \label{fig2}
\end{figure}

Without loss of generality, a Cartesian coordinate system is established according to the red coordinate axes in Fig. 2. The coordinate of source vehicle ${V_t}$ is $\left( {0,0} \right)$, and the coordinate of destination vehicle ${V_d}$ is $\left( { - \frac{{{R_{valid}}}}{{\sqrt 2 }}, - \frac{{{R_{valid}}}}{{\sqrt 2 }}} \right)$. As shown in Fig. 2, $Z$ is the FP if the purple vehicle node is selected as the relay vehicle. ${\rm{\hat C}}\left( {\rm{Z}} \right)$ denotes the green region and ${\rm{\hat D}}\left( {\rm{Z}} \right)$ denotes the yellow region.

\subsection{Forward Progress and Manhattan Distance of the Single Hop Link}
\label{sec3-1}

The PDFs of forward progress and Manhattan distance of the single hop link are given by lemma 1 and 2, respectively.

{\em Lemma 1:} The PDF of FP ${Z_{ran}}$ with random relay selection strategy is given by

\begin{equation}
{f_{{Z_{ran}}}}\left( {{z_{ran}}} \right) = \frac{{\sqrt 2 }}{{{l_t}}},
\end{equation}
where $0 \le {z_{ran}} \le \frac{{{l_t}}}{{\sqrt 2 }}$.

{\em Proof:} FP is uniformly distributed on line ${l_0}$ in Fig. 2 if random relay selection strategy is adopted. The CDF of ${Z_{ran}}$ can be written as ${F_{{Z_{ran}}}}\left( {{z_{ran}}} \right) = P\left( {{Z_{ran}} \le {z_{ran}}} \right) = \frac{{\sqrt 2 {z_{ran}}}}{{{l_t}}}$, where $0 \le {z_{ran}} \le \frac{{{l_t}}}{{\sqrt 2 }}$ . Therefore, the PDF of ${Z_{ran}}$ is ${f_{{Z_{ran}}}}\left( {{z_{ran}}} \right) = \frac{{d{F_{{Z_{ran}}}}\left( {{z_{ran}}} \right)}}{{d{z_{ran}}}} = \frac{{\sqrt 2 }}{{{l_t}}}$, where $0 \le {z_{ran}} \le \frac{{{l_t}}}{{\sqrt 2 }}$.

According to (6), the hop number with random relay selection strategy is derived by

\begin{equation}
\begin{gathered}
  {k_{ran}} = \frac{{{R_{valid}}}}{{\mathbb{E}\left( {{Z_{ran}}} \right)}} \hfill \\
   = \frac{{{R_{valid}}}}{{\int_0^{\frac{{{l_t}}}{{\sqrt 2 }}} {{z_{ran}}{f_{{Z_{ran}}}}\left( {{z_{ran}}} \right)d{z_{ran}}} }} \hfill \\
   = \frac{{2\sqrt 2 {R_{valid}}}}{{{l_t}}} \hfill \\
\end{gathered} .
\end{equation}

{\em Lemma 2:} The PDF of Manhattan distance ${D_{man,ran}}$ with random relay selection strategy is given by

\begin{equation}
{f_{{D_{man,ran}}}}\left( {{d_{man,ran}}} \right) = \frac{{2{d_{man,ran}}}}{{l_t^2}},
\end{equation}
where $0 \leqslant {d_{man,ran}} \leqslant {l_t}$.

{\em Proof:} See Appendix A.

\subsection{SNR of the Single Hop Link}
\label{sec3-2}

For the typical single hop link ${l_{i,j}}$, before the transmitting vehicle ${V_i}$ sends the useful signal to the receiving vehicle ${V_j}$, ${V_i}$ and ${V_j}$ will complete the beam alignments according to the mechanism described in section II-D. Therefore, the transmission and reception directivity antenna gains of vehicles over link ${l_{i,j}}$ are  main lobe gains. The $SNR$ of link ${l_{i,j}}$ can be written as

\begin{equation}
\begin{gathered}
  SN{R_j}[dB] \hfill \\
   = {P_t}[dBm] - {N_0}[{{dBm} \mathord{\left/
 {\vphantom {{dBm} {Hz}}} \right.
 \kern-\nulldelimiterspace} {Hz}}] - 10{\log _{10}}\left( {B\left[ {Hz} \right]} \right) \hfill \\
  {\kern 1pt} {\kern 1pt} {\kern 1pt} {\kern 1pt} {\kern 1pt} {\kern 1pt} {\kern 1pt} {\kern 1pt} {\kern 1pt} {\kern 1pt}  + 2{G_\sphericalangle }[dB] - 10\alpha {\log _{10}}\left( {{d_{man}}} \right) - 69.6 - \rho  \hfill \\
\end{gathered} .
\end{equation}

Since the link ${l_{i,j}}$ is a typical single hop link, for the sake of simplicity, the subscript will be removed in the following text. For example, $SN{R_j}$ will be written as $SNR$.

Assuming $M = {P_t}\left[ {dBm} \right] - {N_0}\left[ {{{dBm} \mathord{\left/
 {\vphantom {{dBm} {Hz}}} \right.
 \kern-\nulldelimiterspace} {Hz}}} \right] - 10{\log _{10}}\left( {B\left[ {Hz} \right]} \right) + 2{G_\sphericalangle }\left[ {dB} \right] - 69.6$, the CDF of $SNR$ with random relay selection strategy is derived by

\begin{equation}
\begin{gathered}
  {F_{SNR,ran}}\left( \gamma  \right) = P\left\{ {SNR \leqslant \gamma } \right\} \hfill \\
   = P\left\{ {M - \gamma  - 10\alpha {{\log }_{10}}\left( {{d_{man,ran}}} \right) \leqslant \rho } \right\} \hfill \\
   = \frac{1}{2} - \frac{1}{2}\mathbb{E}\left[ {erf\left[ {Y\left( \gamma  \right)} \right]} \right] \hfill \\
   = \frac{1}{2} - \frac{1}{2}\int\limits_0^{ + \infty } {erf\left[ {Y\left( \gamma  \right)} \right]{f_{{D_{man,ran}}}}\left( {{d_{man,ran}}} \right)d{d_{man,ran}}}  \hfill \\
   = \frac{1}{2} - \frac{1}{2}\int\limits_0^{{l_t}} {erf\left[ {Y\left( \gamma  \right)} \right]\frac{{2{d_{man,ran}}}}{{l_t^2}}d{d_{man,ran}}}  \hfill \\
\end{gathered} ,
\end{equation}
where $Y\left( \gamma  \right) = \frac{{M - \gamma  - 10\alpha {{\log }_{10}}\left( {{d_{man,ran}}} \right)}}{{\sqrt 2 \sigma }}$ and $erf\left[  \bullet  \right]$ is the error function.

\subsection{The Average Total Delay and Reliability of Message Transmission with Random Relay Selection Strategy}
\label{sec3-3}

According to (7), the average total delay of message transmission with random relay selection strategy is derived by

\begin{equation}
\begin{gathered}
  {T_{multi,ran}} \hfill \\
   = {k_{ran}}\int\limits_0^{ + \infty } {\frac{{{P_S}}}{{\left( {1 - \frac{{{\tau _{i,j}}}}{{{T_t}}}} \right)B{{\log }_2}\left( {1 + \gamma } \right)}}{f_{SNR,ran}}\left( \gamma  \right)d\gamma }  \hfill \\
  {\kern 1pt} {\kern 1pt} {\kern 1pt} {\kern 1pt} {\kern 1pt} {\kern 1pt} {\kern 1pt} {\kern 1pt} {\kern 1pt} {\kern 1pt}  + \left( {{k_{ran}} - 1} \right){T_{proc}} \hfill \\
   = \frac{{2\sqrt 2 {R_{valid}}}}{{{l_t}}}\int\limits_0^{ + \infty } {\frac{{{P_S}}}{{\left( {1 - \frac{{{\tau _{i,j}}}}{{{T_t}}}} \right)B{{\log }_2}\left( {1 + \gamma } \right)}}d{F_{SNR,ran}}\left( \gamma  \right)}  \hfill \\
  {\kern 1pt} {\kern 1pt} {\kern 1pt} {\kern 1pt} {\kern 1pt} {\kern 1pt} {\kern 1pt} {\kern 1pt} {\kern 1pt} {\kern 1pt} {\kern 1pt}  + \left( {\frac{{2\sqrt 2 {R_{valid}}}}{{{l_t}}} - 1} \right){T_{proc}} \hfill \\
\end{gathered} ,
\end{equation}
where ${F_{SNR,ran}}\left( \gamma  \right)$ is given by (14).

According to (9), the average total reliability of message transmission with random relay selection strategy is derived by

\begin{equation}
\begin{gathered}
  {P_{multi,ran}} = {\left[ {1 - {F_{SNR,ran}}\left( \varepsilon  \right)} \right]^{^{{k_{ran}}}}} \hfill \\
   = {\left[ {\frac{1}{2} + \frac{1}{2}\int_0^{{l_t}} {erf\left[ {Y\left( \varepsilon  \right)} \right]\frac{{2{d_{man,ran}}}}{{l_t^2}}d{d_{man,ran}}} } \right]^{\frac{{2\sqrt 2 {R_{valid}}}}{{{l_t}}}}} \hfill \\
\end{gathered} ,
\end{equation}
where $Y\left( \varepsilon  \right) = \frac{{M - \varepsilon  - 10\alpha {{\log }_{10}}\left( {{d_{man,ran}}} \right)}}{{\sqrt 2 \sigma }}$.

\section{Simulation Results and Discussion}
\label{sec4}

In this section, the effects of the path loss exponent and the minimum safe distance between adjacent vehicles on the average total delay and reliability of message transmission are numerically analyzed. Some default parameters are configured as follows: ${R_{valid}} = 500\sqrt 2 m$, ${P_t} = 30{\text{dBm}}$ \cite{14Xiang13}, ${N_0} =  - 174{{{\text{dBm}}} \mathord{\left/
 {\vphantom {{{\text{dBm}}} {{\text{Hz}}}}} \right.
 \kern-\nulldelimiterspace} {{\text{Hz}}}}$ \cite{15Zeng18}, $B = 200M{\text{Hz}}$ \cite{12Ghosh14}, ${R_x} = {R_y} = 50m$, $\mu  = 0.08$ \cite{12Ghosh14}, ${T_t} = 4{\text{ms}}$ \cite{16Xiao11}, ${T_p} = 0.2{\text{ms}}$, ${T_{proc}} = 20\mu s$, ${P_S} = 24000bits$ \cite{173GPP17}, $\psi _i^{{t_x}} = \psi _j^{{r_x}} = {40^{\text{0}}}$ \cite{18Wang09}, $\varphi _{i,j}^{{t_x}} = \varphi _{i,j}^{{r_x}} = {10^{\text{0}}}$, ${G_\sphericalangle } = 10dB$, $\varepsilon  = 5dB$ \cite{19Ge11}, $\sigma  = 4dB$.

\begin{figure}
  \centering
  % Requires \usepackage{graphicx}
  \includegraphics[width=9.5cm,draft=false]{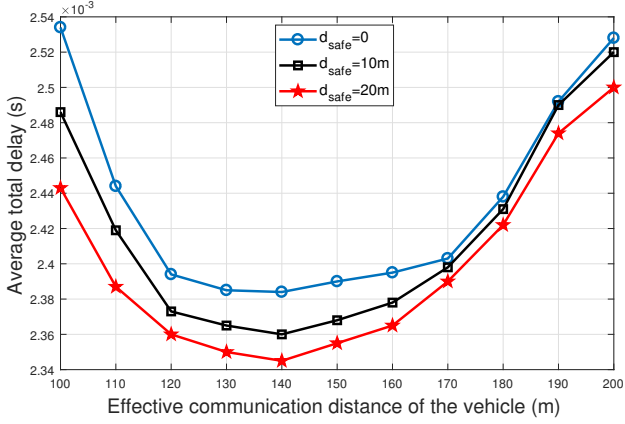}\\
  \caption{\small Average total delay with respect to effective communication distance of the vehicle considering different minimum safe distances ($\alpha  = 2.9$).}
  \label{fig3}
\end{figure}

Fig. 3 shows the average total delay with respect to effective communication distance of each vehicle considering different minimum safe distance. When the minimum safe distance is fixed and the effective communication distance of the vehicle is less than 140 m, the average total delay decreases with the increase of effective communication distance of the vehicle. This result indicates that the average total delay is mainly affected by the hop number in this case. When the minimum safe distance is fixed and the effective communication distance of the vehicle is larger than or equal to 140 m, the average total delay increases with the increase of effective communication distance of the vehicle. This result implies that the average total delay is mainly affected by transmission delay over single hop link in this case. Hence, a minimum average total delay exists with respect to the effective communication distance of the vehicle. When the effective communication distance of the vehicle is fixed, the average total delay decreases with the increase of the minimum safe distance.

\begin{figure}
  \centering
  % Requires \usepackage{graphicx}
  \includegraphics[width=9.5cm,draft=false]{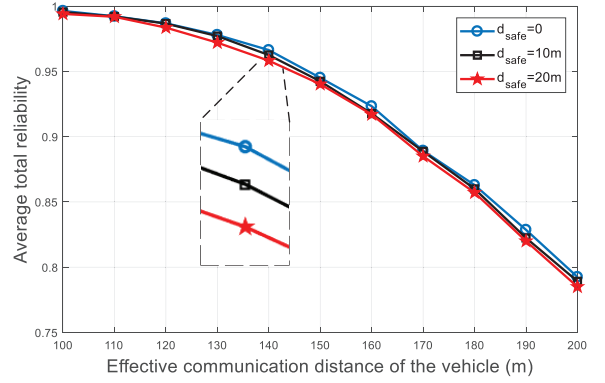}\\
  \caption{\small Average total reliability with respect to effective communication distance of the vehicle considering different minimum safe distances ($\alpha  = 2.9$).}
  \label{fig4}
\end{figure}

Fig. 4 illustrates the average total reliability with respect to effective communication distance of the vehicle considering different minimum safe distances. When the minimum safe distance is fixed, the average total reliability decreases with the increase of effective communication distance of the vehicle.

 \begin{figure}
  \centering
  % Requires \usepackage{graphicx}
  \includegraphics[width=9.5cm,draft=false]{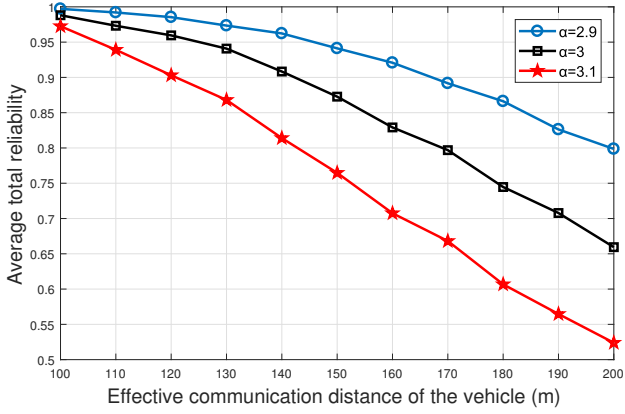}\\
  \caption{\small  Average total reliability with respect to effective communication distance of the vehicle considering different path loss exponents (${d_{safe}} = 4m$).}
  \label{fig5}
\end{figure}

Fig. 5 depicts the average total reliability with respect to effective communication distance of the vehicle considering different path loss exponents. When the path loss exponent is fixed, the average total reliability decreases with the increase of effective communication distance of the vehicle. When the effective communication distance of the vehicle is fixed, the average total reliability decreases with the increase of the path loss exponent. The reason is that path loss becomes severe with the increase of the path loss exponent. Therefore, the channel becomes more unreliable with the increase of the path loss exponent.

\section{Conclusion}
\label{sec5}

In this paper, message transmission delay and reliability between autonomous vehicles based on millimeter wave multi-hop V2V communications  with random relay selection strategy under urban road environment are investigated. Shifted-exponential distribution is proposed to model the headway distance to guarantee a minimum safe distance between adjacent vehicles. Simulation results show that message transmission delay and reliability decrease with the increase of the path loss exponent and are also affected by the minimum safe distance between adjacent vehicles. We will further explore the impact of relay selection strategy on the message transmission delay and reliability in the future.

\section*{Acknowledgment}

The authors would like to acknowledge the support from the National Key Research and Development Program of China under Grant 2017YFE0121600.

\section*{Appendix A}

The CDF of Manhattan distance ${D_{man,ran}}$ with random relay selection strategy is derived by

\begin{equation}
\begin{array}{l}
{F_{{D_{man,ran}}}}\left( {{d_{man,ran}}} \right)\\
 = P\left\{ {{D_{man,ran}} \le {d_{man,ran}}} \right\}\\
 = \frac{{\sum\limits_{{a_n} \in {\rm{\hat D}}\left( {{z_{ran}}} \right)} {{a_n}} }}{{\sum\limits_{{a_n} \in \left\{ {{\rm{\hat C}}\left( {{z_{ran}}} \right) \cup {\rm{\hat D}}\left( {{z_{ran}}} \right)} \right\}} {{a_n}} }}\\
 = \frac{{\eta D\left( {{z_{ran}}} \right)}}{{\eta \left[ {C\left( {{z_{ran}}} \right) + D\left( {{z_{ran}}} \right)} \right]}}\\
 = \frac{{D\left( {{z_{ran}}} \right)}}{{C\left( {{z_{ran}}} \right) + D\left( {{z_{ran}}} \right)}}
\end{array},
\end{equation}
where $0 \le {d_{man,ran}} \le {l_t}$, $\sum\limits_{{a_n} \in {\rm{\hat D}}\left( {{z_{ran}}} \right)} {{a_n}} $ is the road length covered by region ${\rm{\hat D}}\left( {{z_{ran}}} \right)$, $\sum\limits_{{a_n} \in \left\{ {{\rm{\hat C}}\left( {{z_{ran}}} \right) \cup {\rm{\hat D}}\left( {{z_{ran}}} \right)} \right\}} {{a_n}} $ is the road length covered by region ${\rm{\hat C}}\left( {{z_{ran}}} \right)$ and ${\rm{\hat D}}\left( {{z_{ran}}} \right)$, $C\left( {{z_{ran}}} \right)$ is the area of region ${\rm{\hat C}}\left( {{z_{ran}}} \right)$, and $D\left( {{z_{ran}}} \right)$ is the area of region ${\rm{\hat D}}\left( {{z_{ran}}} \right)$.

Obviously, $D\left( {{z_{ran}}} \right) = d_{man,ran}^2$, $C\left( {{z_{ran}}} \right) + D\left( {{z_{ran}}} \right) = l_t^2$. Therefore, the CDF of Manhattan distance ${D_{man,ran}}$ with random relay selection strategy can be written as

\begin{equation}
{F_{{D_{man,ran}}}}\left( {{d_{man,ran}}} \right) = \frac{{d_{man,ran}^2}}{{l_t^2}},
\end{equation}
where $0 \le {d_{man,ran}} \le {l_t}$.

Finally, The PDF of Manhattan distance ${D_{man,ran}}$ with random relay selection strategy can be given by (12).

%\section*{Acknowledgment}
%
% The author would like to acknowledge the support from the NSFC Major International Joint Research Project under the grant 61210002, the Hubei Provincial Science and Technology Department under the grant 2016AHB006, the Fundamental Research Funds for the Central Universities under the grant 2016AHB006 and the support from Graduates¡¯ Innovation Fund, Huazhong University of Science and Technology under Grant 5003181024. This research is partially supported by EU FP7-PEOPLE-IRSES, project acronym CROWN (grant no. 610524).

% conference papers do not normally have an appendix

% use section* for acknowledgement

% trigger a \newpage just before the given reference
% number - used to balance the columns on the last page
% adjust value as needed - may need to be readjusted if
% the document is modified later
%\IEEEtriggeratref{8}
% The "triggered" command can be changed if desired:
%\IEEEtriggercmd{\enlargethispage{-5in}}

% references section

% can use a bibliography generated by BibTeX as a .bbl file
% BibTeX documentation can be easily obtained at:
% http://www.ctan.org/tex-archive/biblio/bibtex/contrib/doc/
% The IEEEtran BibTeX style support page is at:
% http://www.michaelshell.org/tex/ieeetran/bibtex/
%\bibliographystyle{IEEEtran}
% argument is your BibTeX string definitions and bibliography database(s)
%\bibliography{IEEEabrv,../bib/paper}
%
% <OR> manually copy in the resultant .bbl file
% set second argument of \begin to the number of references
% (used to reserve space for the reference number labels box)

% that's all folks
\end{document}